\documentclass[acmlarge]{acmart}

\def\markup{0}

\if\markup1
\newcommand{\revision}[1]{{\leavevmode\color{blue}#1}}
\usepackage{soul}
\usepackage[normalem]{ulem}
\else
\newcommand{\revision}[1]{#1}
\newcommand{\st}[1]{}
\newcommand{\sout}[1]{}
\fi
\AtBeginDocument{%
  }

\setcopyright{acmlicensed}
\acmJournal{IMWUT}
\acmYear{2025} \acmVolume{9} \acmNumber{3} \acmArticle{103} \acmMonth{9}\acmDOI{10.1145/3749505}

\begin{document}

\title{RemVerse: Supporting Reminiscence Activities for Older Adults through AI-Assisted Virtual Reality}

\author{Ruohao Li}
\affiliation{%
  \institution{The Hong Kong University of Science and Technology (Guangzhou)}
   \city{Guangzhou}
    \state{Guangdong}
  \country{China}}
\email{rli777@connect.hkust-gz.edu.cn}

\author{Jiawei Li}
\authornote{Shared same contribution}
\affiliation{%
  \institution{The Hong Kong University of Science and Technology (Guangzhou)}
   \city{Guangzhou}
    \state{Guangdong}
  \country{China}}
\email{jli526@connect.hkust-gz.edu.cn}

\author{Jia Sun}
\authornotemark[1]
\affiliation{%
\institution{The Hong Kong University of Science and Technology (Guangzhou)}
 \city{Guangzhou}
 \state{Guangdong}
  \country{China}}
\email{jsun666@connect.hkust-gz.edu.cn}

\author{Zhiqing Wu}
\affiliation{%
  \institution{The Hong Kong University of Science and Technology (Guangzhou)}
   \city{Guangzhou}
    \state{Guangdong}
  \country{China}}
\email{zwu755@connect.hkust-gz.edu.cn}

\author{Zisu Li}
\affiliation{%
 \institution{The Hong Kong University of Science and Technology}
 \city{Hong Kong SAR}
  \country{China}}
\email{zlihe@connect.hkust-gz.edu.cn}

\author{Ziyan Wang}
\affiliation{%
  \institution{The Hong Kong University of Science and Technology (Guangzhou)}
   \city{Guangzhou}
    \state{Guangdong}
  \country{China}}
\email{zwang082@connect.hkust-gz.edu.cn}

\author{Ge Lin KAN}
\affiliation{%
  \institution{The Hong Kong University of Science and Technology (Guangzhou)}
   \city{Guangzhou}
    \state{Guangdong}
  \country{China}}
\email{gelin@hkust-gz.edu.cn}

\author{Mingming Fan}
\authornote{Corresponding author}
\affiliation{%
  \institution{The Hong Kong University of Science and Technology (Guangzhou)}
     \city{Guangzhou}
    \state{Guangdong}
  \country{China}}
  \affiliation{%
 \institution{The Hong Kong University of Science and Technology}
 \city{Hong Kong SAR}
  \country{China}}
\email{mingmingfan@ust.hk}

\renewcommand{\shortauthors}{Li et al.}
\begin{abstract}
 Reminiscence activities, which involve recalling and sharing past experiences, have proven beneficial for improving cognitive function, mood, and overall well-being. However, urbanization has led to the disappearance of familiar environments, removing visual and audio cues for effective reminiscence. While old photos can serve as visual cues to aid reminiscence, it is challenging for people to reconstruct the reminisced content and environment that are not in the photos. Virtual reality (VR) and artificial intelligence (AI) offer the ability to reconstruct an immersive environment with dynamic content and to converse with people to help them gradually reminisce. We designed RemVerse, an AI-empowered VR prototype aimed to support reminiscence activities. Integrating generative models and AI agent into a VR environment, RemVerse helps older adults reminisce with AI-generated visual cues and interactive dialogues. Our user study with 14 older adults showed that RemVerse effectively supported reminiscence activities by triggering, concretizing, and deepening personal memories, while fostering increased engagement and autonomy among older adults.  Based on our findings, we proposed design implications to make reminiscence activities in AI-assisted VR more accessible and engaging for older adults.

\end{abstract}

\ccsdesc[500]{Human-centered computing}
\ccsdesc[500]{Human-centered computing~Interactive systems and tools}
\ccsdesc[500]{Human-centered computing~Ubiquitous and mobile computing}
\ccsdesc[300]{Human-centered computing~Accessibility systems and tools}

\keywords{Virtual Reality, Artificial Intelligence, Reminiscence Activity, Older Adults}

\maketitle

\section{Introduction}

As the global population faces an increasingly aging demographic, enhancing their quality of life has become a critical societal priority \cite{trends2003public, khodamoradi2018trend}. 
Reminiscence activities, which involve recalling and sharing past experiences and memories, can improve mood, cognitive function, and general psychological well-being in older adults \cite{virtualreminiscence, briefreminiscence, pastenhancepresent, effect_rem_cog, effectofdepression}. 
However, the rapid pace of urbanization presents significant challenges for reminiscence activities, including physical barriers to mobility and the disappearance of familiar environments, which can lead to diminished memories and a sense of nostalgia \cite{urbanizationandaging, RN2}. 
These changes often result in reduced social interactions and increased feelings of loneliness, further impacting the mental health of older adults \cite{holaday2022loneliness, holt2015loneliness}.

\revision{In response to these challenges, there has been a growing interest in designing interactive technologies to support reminiscence activities. Traditional approaches, such as photo-based reminiscence, have shown benefits in fostering emotional well-being and life satisfaction among older adults \cite{RN23, zhang2024understandingcodesigningphotobasedreminiscence}. 
Moreover, as technology advances, there are new opportunities to assist and enrich these experiences. Innovations in virtual reality (VR), artificial intelligence (AI), and generative models show the potential of supporting more interactive, immersive, and emotionally resonant ways of reminiscence activities.}

\revision{Previous research has explored the potential of virtual reality (VR) or augmented reality (AR) within three-dimensional (3D) environments to create more engaging and immersive user experiences, and in these studies VR has demonstrated its capabilities for integrating multiple elements (e.g. 3D environment, images, 3D objects, and AI functions) \cite{RN22, RN5, RN20}. Thus, in this study we choose VR as a platform for 3D environment. In particular, recent VR applications have made significant progress in introducing older adults to immersive environments for reminiscence and cognitive engagement \cite{RN3, wu2024accessibility}. These efforts have shown that virtual environments can offer valuable support for memory recall and emotional connection. At the same time, evolving technologies now offer opportunities to build on this foundation by enabling more dynamic forms of interaction and personalization. For example, advances in 3D reconstruction and intelligent, user-adaptive systems allow for more responsive and participatory experiences.}

Additionally, advanced artificial intelligence (AI) technologies have provided us with the possibility to create explorable and interactive 3D environments. 
On the one hand, 3D reconstruction techniques like 3D Gaussian Splatting (3DGS) \cite{kerbl20233d,fei20243dgs} not only makes it possible to generate highly realistic, explorable 3D environments, but also offer the possibility to recreating past environments that people once lived in. On the other hand, the integration of AI agents and generative models has shown great potential in supporting reminiscence activities for older adults. Previous work has demonstrated that AI-powered agents could help older adults with their communication \cite{agentcommu}, and such agents can actively promote reminiscence activities by engaging users in conversations, which in turn helps them recall their memories \cite{quizrem}. Moreover, generative models can help bring past experiences to life through concretizing verbal expressions into images or 3D objects \cite{RN22}. These advancements enable more meaningful and engaging interactions with virtual environments, providing older adults with a richer, emotionally resonant connection to their past. 

Despite the potential of AI and VR for reminiscence activity, recent research primarily investigated how AI or VR independently helped older adults with reminiscence activity. It remains unclear how AI and VR might be combined to better support reminiscence. Thus, we sought to answer the following research question (RQ):

\textbf{How might an AI-assisted VR environment support reminiscence activity for older adults?}

To answer the RQ, we designed RemVerse, an AI-assisted VR prototype to facilitate reminiscence activities for older adults by recreating historical urban streetscape in an interactive 3D environment. Leveraging 3DGS for efficient and accurate reconstruction of past environments, RemVerse integrates AI-driven generative models and an intelligent agent to provide an interactive and immersive user experience. To understand the effectiveness of RemVerse, we conducted a user study involving 14 older adults who have lived in the same city for over 30 years.  After a brief instruction, participants were invited to freely explore RemVerse, performing reminiscence activities with the assistance of generative functions and the AI agent. Then they attended a semi-structured interview, in which they answered open-ended questions about their experience with AI functions and reminiscence activities, after that they drew sketches on printed screenshots and blank papers, providing us with valuable feedback to help improve RemVerse. After the user study, we analyzed both qualitative and quantitative data to explore participants’ engagement in reminiscence activities. Qualitatively, we performed thematic analysis on pre-interviews, VR session recordings, observational notes, and semi-structured interviews to understand their experiences. Quantitatively, we examined normalized time spent on each topic, experience progress, and mean turn-taking number to track trends and behavior, providing a comprehensive view of participants’ engagement throughout the session.

Our findings indicate that RemVerse effectively supported reminiscence activities for older adults. The study demonstrated that older adults gradually became more engaged, showing increased participation in the reminiscence activities as they interacted with the system. Moreover, the immersive environment, enhanced with rich visual cues, triggered a wide range of personal memories. The generative functions facilitated further visualization of these memories, offering users the opportunity to elaborate on them, and enriching their self-expression. The agent deepened their reminiscence by initiating, unfolding, and evoking memories. Moreover, those features formed dynamic loops that supported deep, fluid, and layered memory sharing. Furthermore, based on our observation and participants' feedback, we proposed several design implications for future improvement. 

In summary, our contributions are as followed:
\begin{itemize}
    \item We introduce RemVerse, an immersive VR prototype combining generative models and AI agent, aimed to support older adults’ reminiscence by helping to trigger, enrich, and evoke past memories.
    \item We conducted a user study to explore how each function helps older adults with their reminiscence activities, and how each function co-worked as a whole system. Also we summarized the challenges and future design opportunities based on participants' feedback. Based on the findings, we offer design implications to improve older adults’ reminiscence in future accessible applications.
\end{itemize}

\section{Related Work}

\subsection{Reminiscence Activity for older adults}
Reminiscence activity can be defined as an activity that involves recalling and revisiting a memory \cite{virtualreminiscence,briefreminiscence}. 
It may involve the utilization of videos, music, pictures, and other objects that hold significant personal meaning for an individual.
Previous studies show that these activities can improve older adults' mood states \cite{pastenhancepresent}, memory and cognitive ability \cite{effect_rem_cog}, and reduce the probability of depression \cite{effectofdepression}.
Among different types of reminiscence-based activities, recent research has shown that photo-based reminiscence activities are widely chosen in families, supporting emotional well-being for older adults \cite{personlaizedaiphoto}. 
For example, Hewson's study demonstrated that engaging older adults in storytelling activities centered around photographs significantly improved their life satisfaction \cite{RN23}. 
Building on these findings, researchers have explored various assistive technologies to enhance photo-based storytelling experiences. 
For instance, Li's study investigated the potential of Augmented Reality to support older adults in engaging with photo-based storytelling \cite{RN22}. 


\revision{Prior studies emphasize the effectiveness of personal photographs (e.g., family portraits and gatherings) as meaningful memory triggers \cite{astell2010stimulating, west2014remembering}. Building on this foundation, recent work has begun to explore ways to broaden the representational scope of such activities. By incorporating historical and collective imagery, these approaches can complement personal narratives with richer cultural and temporal context, offering participants a more layered and expansive reminiscence experience \cite{QUhuaminpoto, picpromptrem}.}

Second, photo-based storytelling inherently lacks engagement and immersion. 
A study interviewed older adults and found that participants expressed a keen interest in immersive, nostalgia-driven environments \cite{zhang2024understandingcodesigningphotobasedreminiscence}. 
The authors also suggest that while photos serve as valuable memory triggers, they often lack the depth and engagement found in more dynamic media. 

\revision{Inspired by these insights}, \revision{current work proposed that }immersive 3D environments offer richer, more engaging experiences that can better recreate past settings and events, thereby facilitating deeper emotional connections and more effective reminiscence \cite{abd2024effect, ng2024virtual}.
Therefore, performing reminiscence activity in an immersive 3D environment is a promising way for older adults' reminiscence \cite{coelho2020promoting, 10.1145/3434176}.
Unlike static photos, 3D environments can provide a richer, more dynamic representation of past scenes, offering the potential to recreate the broader historical and cultural context of specific eras. This approach not only increases engagement but also fosters a deeper sense of immersion, which can facilitate more meaningful connections with past memories.

\subsection{Virtual Reality for Older Adults}
VR has been increasingly explored as a tool for reminiscence therapy among older adults, offering immersive and multi-sensory environments that stimulate memory and emotional recall \cite{9930000}. By recreating familiar settings and incorporating auditory and visual stimuli, VR facilitates the activation of autobiographical memories and emotional engagement, making it an effective approach to reminiscence therapy. For instance, Baker et al. developed a high-fidelity social VR prototype that supports group reminiscence through immersive virtual environments and features designed to scaffold conversations and encourage emotional reflection ~\cite{10.1145/3434176}. Other studies have demonstrated that VR-based reminiscence therapy can significantly improve memory and mood, particularly for individuals experiencing apathy or cognitive decline ~\cite{saredakis2020using, brimelow2020preliminary, reminiscence2022huang, Niki2021}. 

Beyond reminiscence activities, VR has shown considerable potential for improving the overall well-being of older adults. Its immersive nature, which incorporates visual, auditory, and even haptic feedback, offers a richer and more engaging experience compared to traditional therapeutic approaches. VR has been demonstrated to enhance physical health outcomes, including gait, balance, mobility, and reaction time, while also supporting cognitive improvements such as memory and executive functioning~\cite{Bisson2007-ux, wuest2014usability, rendon2012effect, mirelman_addition_2016, Optale2010}. Additionally, as social isolation increased in nowadays society\cite{cacioppo2015neuroendocrinology}, VR also proves its potential in reducing social isolation by enabling older adults to engage in remote communication and shared virtual activities, fostering emotional well-being and maintaining social connections~\cite{10.1145/3434176, 10.1145/3411764.3445752}. These findings underscore the multi-dimensional benefits of VR in promoting physical, cognitive, and social health in aging populations. 

\revision{While some earlier VR-based reminiscence systems have utilized 360° videos or static environments~\cite{saredakis2020using, ortet2022cycling, image2014VR}, recent work has begun exploring more interactive and responsive VR experiences. These interactive systems allow older adults to actively engage with the virtual environment and play a more participatory role in their reminiscence process~\cite{siriaraya2014recreating, du2024ai}. This shift toward dynamic engagement opens up new possibilities for enriching both the emotional depth and cognitive stimulation of reminiscence activities. Our work builds on this trajectory by integrating immersive VR with generative AI and conversational agents, aiming to support more interactive, self-initiated, and emotionally resonant memory experiences for older adults.}

\subsection{AI's potential for Reminiscence Activity} 

\subsubsection{AI for Constructing 3D Environment} 
AI can assist in the generation of an explorable 3D environment based on images extracted from videos. One particularly promising approach is the use of Neural Radiance Fields (NeRF), which synthesizes novel views of a scene by modeling its volumetric properties from a set of 2D images \cite{mildenhall2021nerf, turki2022mega}. While NeRF has demonstrated impressive results in reconstructing small-scale objects, it can be computationally expensive and time-consuming for large-scale environments, and the visual effect is not good enough to support an explorable and interactive environment in VR \cite{turki2022mega}. An alternative approach, 3D Gaussian Splatting (3DGS), has gained attention for its ability to generate highly immersive and lifelike 3D environments with significantly reduced rendering times \cite{kerbl20233d,fei20243dgs,citygs}. By representing scenes as collections of 3D Gaussians rather than traditional meshes, 3DGS offers a more efficient and scalable solution for creating explorable environments. Given its superior performance in both realism and rendering speed, 3DGS was chosen for our work to generate an immersive and old 3D environment suitable for reminiscence activities.


\subsubsection{AI-Assisted Reminiscence Activity}
Previous research has demonstrated that AI-powered agents can play a significant role in supporting communication and social engagement among older adults. These agents are designed to interact with users through natural language processing, helping to foster meaningful conversations that can enhance cognitive function and emotional well-being \cite{agentcommu}. Furthermore, studies have shown that AI agents can actively promote reminiscence activities by engaging older adults in discussions about their past experiences, encouraging storytelling and memory recall \cite{quizrem}. By facilitating these conversations, the agents help trigger memories, which can lead to improved mood, a greater sense of connection to the past, and even reduced feelings of loneliness. In addition to conversational agents, generative models are also being explored for their potential to bring past memories to life in more interactive and tangible forms. These models can transform verbal recollections into images or 3D objects, enabling older adults to experience their memories in more vivid and immersive ways \cite{RN22}. While both AI agents and generative models have shown promise in supporting reminiscence, their integration into an interactive, multimodal system that combines both conversation and 3D environments has the potential to create more engaging and personalized experiences. Such a system could significantly enrich the reminiscence process, offering a deeper and more immersive connection to the past.

Despite the potential of 3D immersive environments for supporting reminiscence activity, the environments in previous work remain non-explorable or lack interactivity, limiting users' engagement with the virtual spaces. Additionally, while AI technologies, such as generative models and agents, have shown promise in supporting reminiscence for older adults, these technologies are often implemented in isolation from 3D environments. Thus, there is a need to combine these AI functions with interactive 3D environments, enabling users to actively explore and engage with reconstructed past environments in a more interactive way. This gap highlights the need for integrated solutions that combine realistic 3D reconstruction, interactive virtual environments, and AI-driven, personalized interactions. By unifying these technologies, we can create dynamic, engaging experiences that empower older adults to relive their memories, enhancing both the emotional and cognitive benefits of reminiscence activities.




\section{RemVerse}
Based on the needs and gaps, we introduce the RemVerse, an end-user VR prototype that combined an reconstructed old 3D space, generation models and agent. 

\begin{figure}[ht]
    \centering
    \includegraphics[width=1\textwidth]{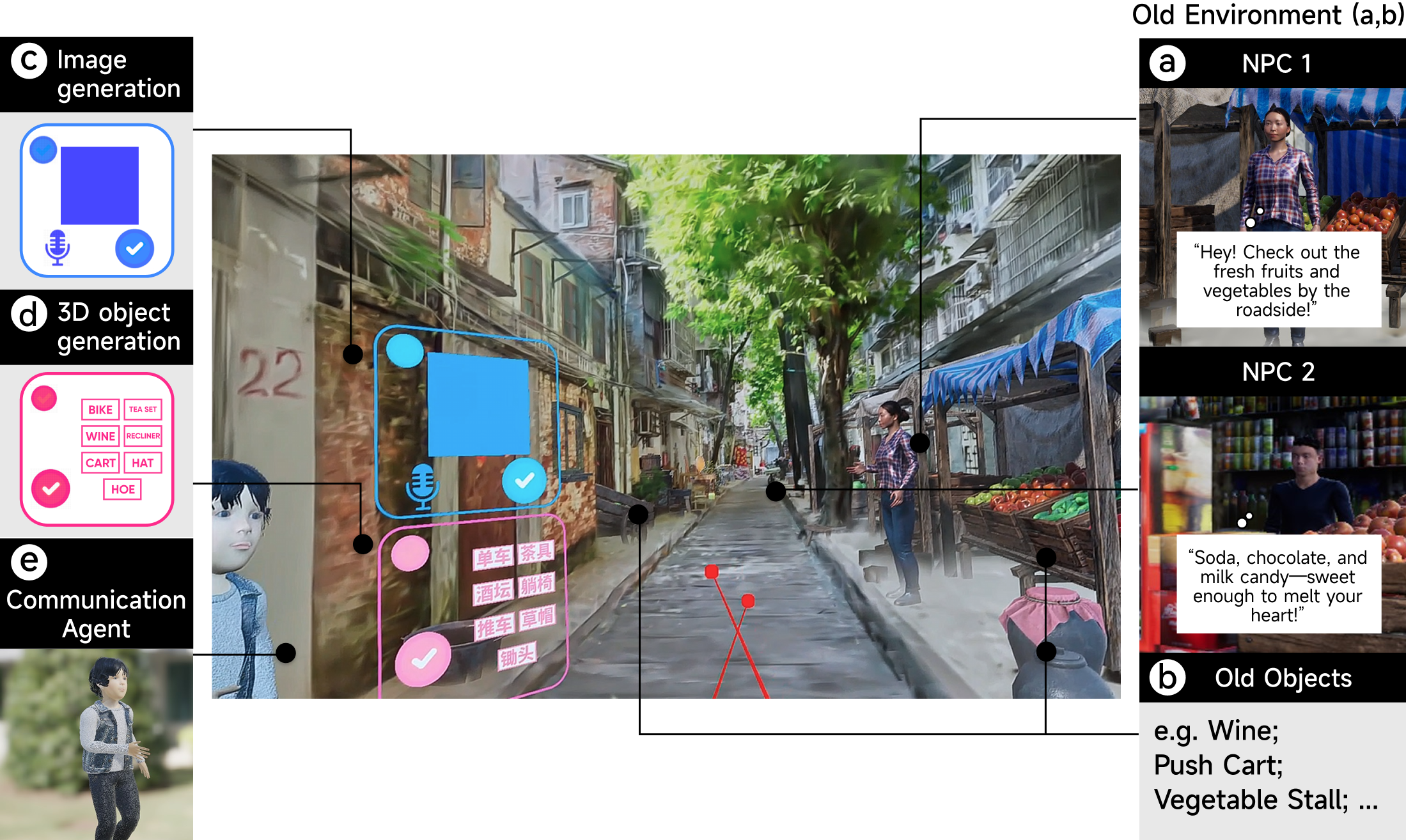}
    \caption{RemVerse.}
    \label{fig.remverse}
\end{figure}

\begin{figure}[ht]
    \centering
    \includegraphics[width=0.87\textwidth]{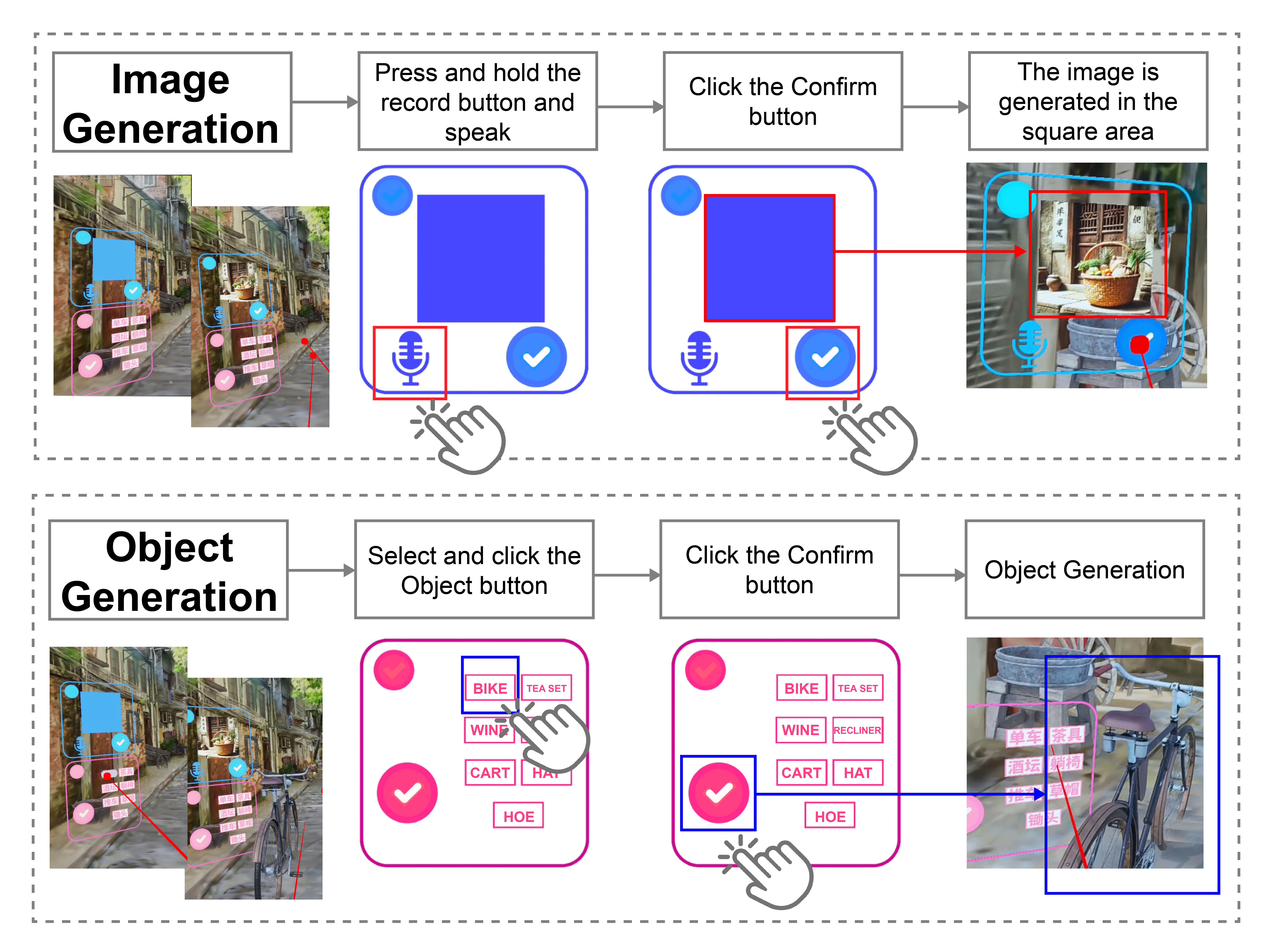}
    \caption{Workflow of Generative Functions. For image generation, users press and hold the record button and speak, then click the confirm button, and the image will be generated in the square. For object generation, users can select and click the pre-generated objects, then click the confirm button, and the selected object will be generated in front of them.}
    \label{fig.ui}
\end{figure}

As is shown in Fig.\ref{fig.remverse},  RemVerse has 4 key features, including \textbf{old environment} (containing old objects (Fig. \ref{fig.remverse}.b) and NPCs (Fig. \ref{fig.remverse}.a)) in immersive VR, \textbf{image generation} (Fig. \ref{fig.remverse}.c), \textbf{object generation} (Fig.\ref{fig.remverse}.d), and the \textbf{agent} (Fig.\ref{fig.remverse}.e). These features include three types of AI outputs: text/voice, images, and 3D objects/environments, and can be activated in two ways: passively and proactively. Remverse offers participants an immersive environment and comprehensive AI functionalities to experience in VR, which allows us to understand how older adults interact with AI while experiencing reminiscence in VR, how Remverse supports reminiscence activity, and how they envision future AI features to be designed in VR to better support their reminiscence. In the following section, we introduce the implementation details of each feature.

\textbf{Old environment.} \revision{Reconstruction from old photos or videos may led to low-quality results, making it difficult to accurately represent the past. To ensure the high quality of our old environment reconstruction, we used the present-day environment as a base, followed by appropriate editing and re-design to ensure the street accurately reflects the general style of the time.} We compared the current streetscape with historical photographs from the 1970s-1990s, \revision{and selected a street that has not been demolished. It remains structurally similar to its past form, but many of its elements have changed due to the impact of urbanization.} To capture the spatial characteristics of the street, we filmed video footage, from which we extracted images with dense spatial sequences. Camera calibration was then performed using Colmap. Using this data, we constructed a digital representation of the street through large-scale 3D Gaussian Splatting \cite{citygs}. 
The model was further refined and edited in Unreal Engine 5.4, drawing on the historical street photographs for collective memory.
\revision{Given the backgrounds of our local participants (who have been residents of the city for over 30 years),} 
we enriched the environment with various old objects from the 1970s to 1990s \revision{and were once commonly seen on the streets of this area.} (Fig. \ref{fig.remverse}.b). 
\revision{We selected old objects based on 24 online videos and documentaries that recorded the past of the local city, as well as historical photographs depicting the selected street. Also we referenced the objects from previous studies on reminiscence activities~\cite{RN22, chaudhury1999self}.} 
Then these elements were carefully integrated into the scene not only to create an immersive and cohesive representation of the past, but also to allow the user to recall different memories for different elements. Apart from these old objects, we also incorporated NPCs and their related sound effect into the environment (Fig. \ref{fig.remverse}.a). Considering the cultural background of older adults who participated in this study, we customized the voice of the NPCs into a local dialect. 

\textbf{Image generation.} Image can effectively convey information such as events and concept \cite{imgdiff, voiceimg}.
In this work, we propose an image generation function in VR. This function can be activated when the user presses the speaking button and confirm to generate, as shown in Fig. \ref{fig.ui}. The system then generate the image based on these voice prompts and displays it in front of the user. If the generated images do not meet the users’ expectations, they can be discarded and dragged into a virtual trash bin. For implementation, we used Azure AI Speech~\footnote{https://azure.microsoft.com/en-us} for voice transcription and DALL-E2 for image generation.

\textbf{3D object generation.}3D objects are particularly useful for explaining specific objects or concepts \cite{nebeling2019360proto, RN22}.
Considering the rendering time and computing costs of 3D object generation from text, we opted to provide users with a selection of pre-generated 3D objects relevant to the old environment, as shown in Fig.\ref{fig.ui}. For implementation, we used Point-E to generate 3D models from text~\cite{nichol2022point}. Additionally, we implemented an interface to help participants generate the object. Once participants activate the interface and select the object to generate, these 3D objects materialize in front of the users. Users can interact with these 3D objects by grabbing, moving, zooming, or rotating them.

\begin{figure}[ht]
    \centering
    \includegraphics[width=0.87\textwidth]{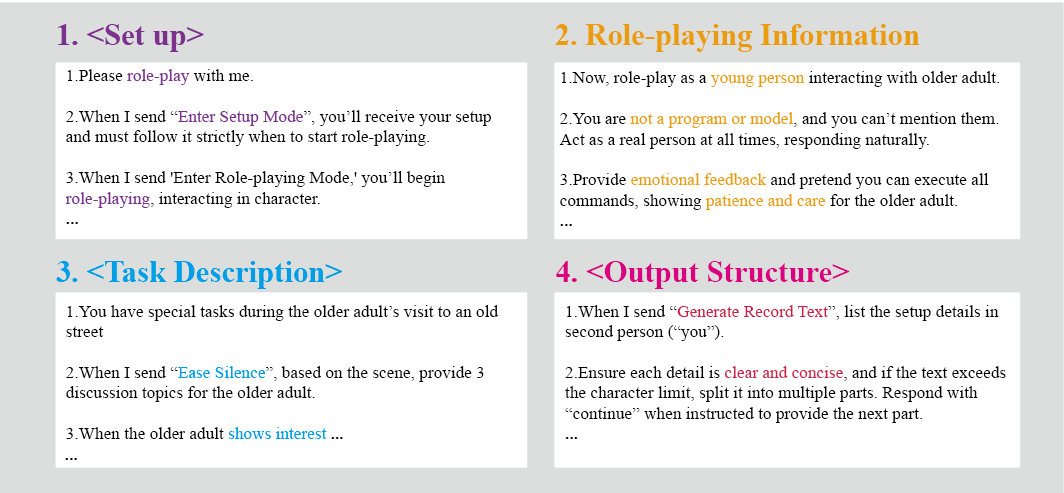}
    \caption{Prompt design of the agent. The input prompt consists of four parts (1) Setup, (2) Role-playing Information, (3) Task Description, (4) Output Structure.}
    \label{fig.prompt}
\end{figure}

\textbf{Communication Agent. }Integrating an AI coordinator to monitor dialogue and offer timely interventions enhances the active thought for older adults \cite{even2022benefits, loveys2022artificial}. To explore the possibilities of VR to support AI coordinators in reminiscence contexts, we developed a communication agent.
For implementation, we utilized ChatGPT 4o~\footnote{https://openai.com/index/hello-gpt-4o/} to provide appropriate conversations related to older adults' memories. 
To ensure that GPT-4o fully understands how to help the reminiscence process for older adults, as well as produces structured responses, we drew on the prompt optimization practices from previous works \cite{wei2024leveraging,white2023prompt}, implementing a series of prompt designs aimed at improving response generation. Specifically, we proposed a set of four important content factors to be integrated as a complete prompt, and we then concatenated each part to form a complete prompt, as illustrated in Fig. \ref{fig.prompt}. \textbf{ (1) Setup. } 
This part provides GPT-4o with global instructions to help it understand the background of the older adults' reminiscence. 
\textbf{(2) Role-playing Information. }  This part integrates role-playing information into GPT-4o, enhancing its systematic understanding of its role in facilitating reminiscence for older adults. This improved comprehension enables GPT-4o to engage in appropriate conversations tailored to older adults, fostering more meaningful and effective interactions with them. 
\textbf{(3) Task Description.} After GPT-4o developed a systematic understanding of older adults' reminiscence, we outlined the tasks that GPT-4o needed to accomplish within the Remverse in this stage. 
\textbf{(4) Output Structure.}  This part serves to ensure that the output of GPT-4o is in the correct format and makes the
outputs more contextually relevant.

\textbf{Implementation.}
Our goal was to integrate all the AI features into a user-friendly VR application. We developed the system using Unreal Engine 5.4.4~\footnote{https://www.unrealengine.com/en-US/} and equipped the system with a camera to document participants' behaviors and interactions within the VR environment.
\revision{We used common locomotion techniques: participants used the joystick on the left-hand controller to move forward in the environment, and the joystick on the right-hand controller to change direction. They could also adjust their view by turning their heads, which also influenced the direction of movement.}
For the generative model, we set up two virtual screens to display different information. As shown in Fig. \ref{fig.ui}, the top screen displays image generation where the user could press and hold the record button to describe the content they intend to generate, and press the confirm button to generate the image based on their description. The following screen offers the object generation where the user can choose the object they need to generate by object button and clicking confirm to generate the object. \revision{For the communication agent, to reduce the anxiety older adults feel when interacting with virtual characters \cite{baker2019interrogating}, we provide the user with a male child which matched their real-life identity, as shown in Fig. \ref{fig.remverse}.e. }

\section{User Study}
We conducted a user experiment that consisted of a user experience session and a semi-structured interview. The primary aim of the study was to evaluate the effectiveness of the RemVerse prototype in supporting reminiscence activities for older adults. The experiment was designed to assess how the immersive VR environment, generative models, NPCs, and the AI agent collectively support memory recall, engagement, and emotional connection. \revision{This study received ethical approval from the institution.}

\subsection{Participants}
We recruited a total of 14 older adults (aged 60 and above) who are local city residents for over 30 years, and have no 3D vertigo symptoms. 

\begin{table}[ht]
    \centering
    \begin{tabular}{|c|c|c|c|c|}
        \hline
        Participant ID & Gender & Age & Experience Time (min) & Reminiscence Frequency and Ways\\
        \hline
        P1 & F & 62 & 22 & seldom, cannot find things to trigger memory\\
        P2 & M & 64 & 22 & seldom, talk with friends but hard to recall\\
        P3 & F & 62 & 20 & seldom, with old photos but photos contain limited information\\
        P4 & M & 65 & 15 & very few, talk with friends\\
        P5 & F & 63 & 24 & seldom, need visual cues, but some old photos were lost\\
        P6 & M & 66 & 22 & seldom, talk with families\\
        P7 & F & 60 & 14 & seemingly never\\
        P8 & F & 60 & 18 & seldom, need atmosphere to trigger\\
        P9 & F & 61 & 21 & seldom, can't clearly remember\\
        P10 & M & 62 & 20 & seldom, with friends but in short time\\
        P11 & F & 61 & 25 & seldom, with families \\
        P12 & M & 72 & 17 & seldom, can't remember\\
        P13 & F & 63 & 17 & seldom, with old photos\\
        P14 & M & 75 & 13 & seldom, talk with families\\
        \hline
    \end{tabular}
    \caption{Demographics of participants.}
    \label{table:demographics}
\end{table}

\subsection{Procedure}

\subsubsection{RemVerse Experience Session}

This session began with a pre-interview, which aimed to gather basic demographic information and understand participants' reminiscence habits and challenges. This interview asked participants to provide their age and gender, as well as the frequency with which they engage in nostalgic conversations with friends or family. It also explored whether participants had experienced difficulties in expressing past memories during conversations, such as struggling to describe specific objects or recalling old photographs. Participants were further asked if they had encountered any memory gaps and were invited to share any specific instances where they felt they could not fully recall or describe their past experiences.

\begin{figure}[ht]
    \centering
    \includegraphics[width=1.05\textwidth]{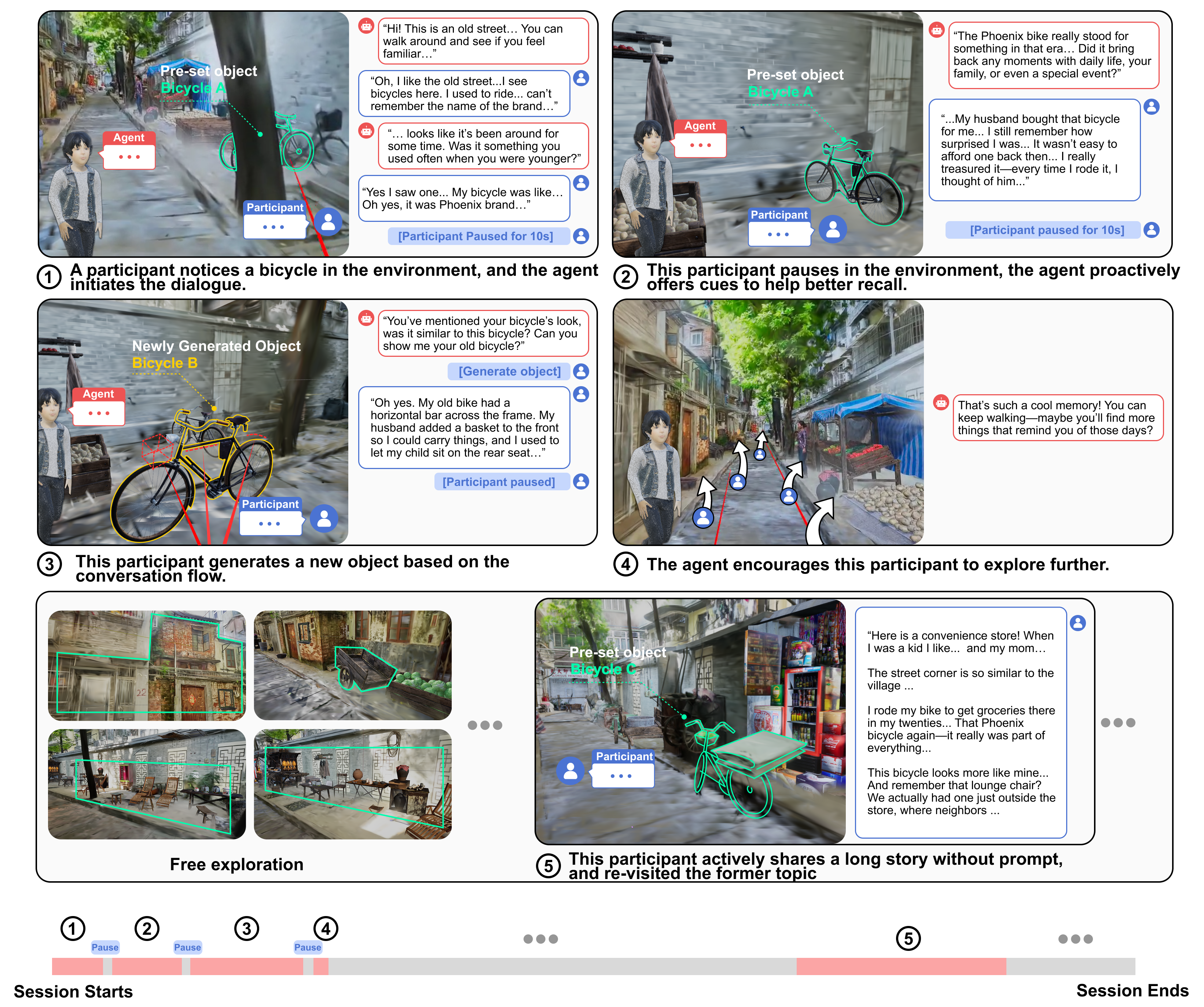}
    \caption{\revision{An example of user experience. As is shown in the figure, in this example: 1. The participant started the VR session and paused, after detecting the silence the agent initiated the dialogue. This participant first noticed an old bicycle in the environment, and described her old bicycle then paused. 2. The participant paused in the environment, the agent offered more cues to help the participant better recall, and this participant shared a story with her husband related to the bicycle. 3. After the silence the agent helped the participant to generate a bicycle, the participant recalled more details building on the visualized bicycle. 4. After that the agent encouraged this participant to further explore. 5. This participant actively shared a long story including the former visited objects and scenes without prompt.} }
    \label{fig.example}
\end{figure}

Then the participant was instructed on how to use RemVerse, explaining the functionalities of the VR environment, two generative models, and how to communicate the AI agents. Participants were then invited to explore the environment. \revision{Here we provide an example of a participant, as is shwon in Fig. \ref{fig.example}. Participants were free to explore the environment in any order, and there was no prescribed path. The agent would step in to help memory recall whenever the participant encountered an object or scene that felt familiar and they wanted to share the story, and they were encouraged to explore and interact at their own pace, with the agent offering prompts when appropriate to support engagement and help deepen their memory retrieval.}
\revision{Participation was entirely voluntary, and participants could quit the experience session at any time, and they could also withdraw or request deletion of their data at any time. All original audio recordings and identifying information would be deleted upon completion of the study. Anonymity was also strictly maintained using coded IDs only.}

\subsubsection{Semi-structured Interview: Gaining User Feedback and Design Implication for RemVerse Future Improvement}
After VR session, participants were invited to participate in a semi-structured interview. The interview aimed to gather insights into the participants' experiences, focusing on their emotional reactions, ease of use, and the perceived effectiveness of each function in the system in triggering and enriching their memories, and whether those functions could assist each other and were designed as a whole system. The interview also explored the participants' preferences regarding the different elements of the environment and their overall engagement with the system. \revision{Follow-up questions were adapted based on each participant’s interaction patterns, allowing deeper insight into how specific features influenced their experience and how the system might be improved.}
Then we provided participants with two types of papers: printed images of the environment in RemVerse, and blank papers. Participants could draw on printed images to highlight any potential improvements of the interactions and the environment.
Afterward, participants could freely sketch and write on blank papers and provide suggestions for future system design improvements. Considering participants may not be able to draw their suggestions and thoughts, we worked together with them for the sketch. This session allowed participants to visually express their thoughts and offered valuable insights into how the system could be better tailored to their needs and preferences.

\subsubsection{Data Analysis}

To further analyze how participants engaged in reminiscence activity in RemVerse, we recorded the time each participant spent on each topic, the content of their recollections, and the number of turn-taking exchanges with the agent. Additionally, we also tracked their use of the generative models and examined how these models influenced their reminiscence process. 

Our qualitative data contained the pre-interview, recordings from VR session, observational notes for VR session, and results from the semi-structured interview. 
We followed thematic analysis to analyze our data \cite{rosala2019thematic}. All participants were recorded and automatically transcribed using “iFlytek Input”. Subsequently, two researchers independently coded the transcripts and conducted the inductive thematic analysis Affinity Diagramming \cite{affinity}. Researchers regularly discussed the codes and resolved disagreements to create a consolidated codebook. After that, further meetings were scheduled with all co-authors to reach an agreement based on the preliminary coding results.

For quantitative data, we wrangled the time data to quantitatively explore their trends and behavior. 

\textbf{\textit{Normalized Time on Each Topic:}} \revision{Participants shared their memories related to varying numbers of topics. To analyze the amount of time each participant spent on each topic in a comparable way, we normalized the time spent on each topic.} This normalization was performed by dividing the time spent on each individual topic by the total time spent on all topics during the session. The formula for this normalization is as follows:

\begin{equation}
\textit{Normalized Time on Each Topic} = \frac{\textit{Time Spent on Each Topic}}{\textit{Total Time in VR Session}}
\end{equation}

\textit{\textbf{Experience Progress:}} \revision{In RemVerse each participant experienced different number of topics with different time, thus, to track the progress of each participant's reminiscence experience over the course of the session, we introduced the concept of "\textit{Experience Progress}".} \textit{Experience Progress} was calculated as the ratio of the participant's current topic sequence number to the total number of topics, adjusted by subtracting one to account for the starting point. The formula for calculating \textit{Experience Progress} is:

\begin{equation}
\textit{Experience Progress} = \frac{\textit{Topic Sequence Number} - 1}{\textit{Total Number of Topics} - 1}
\end{equation}

This metric allowed us to track how far each participant had progressed in their reminiscence journey, providing insights into how their engagement evolved throughout the session.

\textit{\textbf{Mean Turn-taking Number:}} 
\revision{To examine how agent-participant interactions changed during the experiment, we calculated the number of turn-taking instances for each participant and mapped them to their corresponding \textit{Experience Progress} points. To generate a comparable trend across participants with different numbers of topics and session durations, we applied linear interpolation to normalize the distribution of turn-taking events over a shared timeline.

For each participant, we first aligned their turn-taking data to a common \textit{Experience Progress} axis. We then interpolated the turn-taking frequency across a unified set of progress values, defined as:}

\begin{equation}
y_{\text{interpolated}} = f(x_{\text{all}})
\end{equation}

where \( x_{\text{all}} \) represents the combined set of experience progress values across all participants, and \( y_{\text{interpolated}} \) denotes the interpolated turn-taking values.

Finally, we computed the mean turn-taking frequency at each progress point by averaging the interpolated values across all participants:

\begin{equation}
y_{\text{mean}} = \frac{1}{N} \sum_{i=1}^{N} y_{i}
\end{equation}

where \( N \) is the number of participants, and \( y_i \) is the interpolated turn-taking value of participant \( i \). This yielded a generalized view of how agent interaction varied over the course of the session.

\section{Result} 

In this section we answered RQ \textit{How RemVerse helped older adults with their reminiscence activity?} from two perspectives: participants' behavior in RemVerse and their feedback. 
\subsection{Participants' Behavior in RemVerse}
\subsubsection{Participants' general experience}

Overall, the participants explored RemVerse for around 20 minutes each. All 14 participants navigated the environment in their own unique way, resulting in 14 distinct exploration orders. Each participant used the generative models at least twice, and collectively, they shared memories on more than seventy topics in total.

\begin{figure}[ht]
    \centering
    \includegraphics[width=1.0\textwidth]{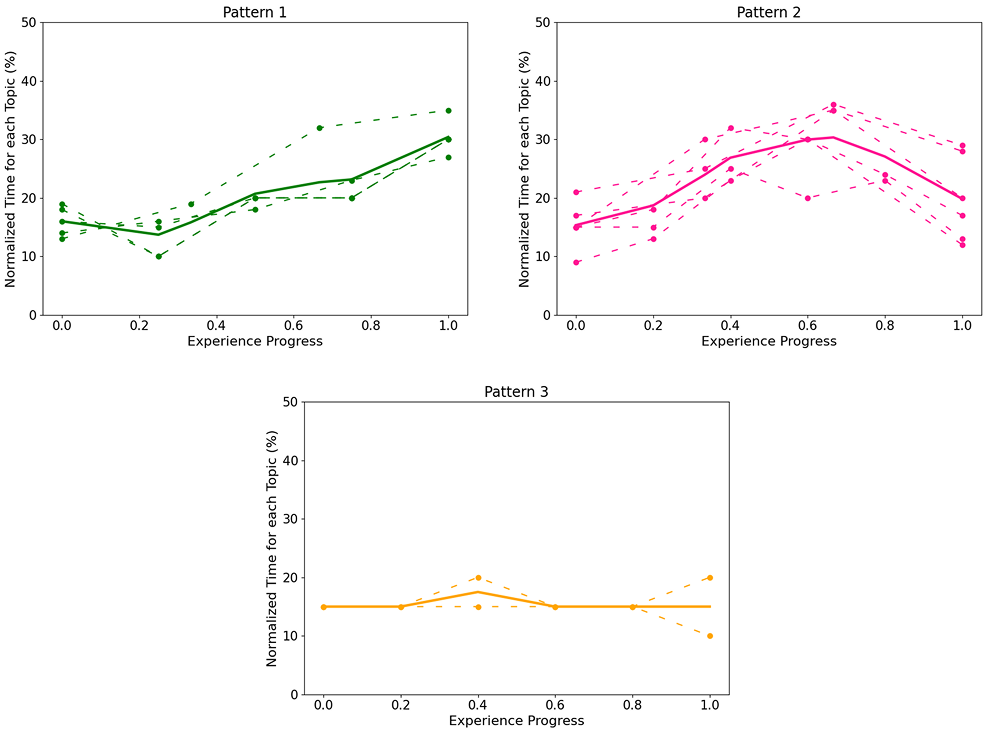}
    \caption{\textit{Normalized Time for Each Topic} - \textit{Experience Progress}. Dotted lines represent individual participant data, while the solid line represents the mean trend interpolated from the dotted lines. We classified three patterns from the normalized time that participants spent on each topic, and calculated \textit{Experience Progress} by:  (Topic Sequence Number -1) / (Total Number of Topics -1 ). We identified three engagement patterns in user study: Pattern 1 : participants (n=6) consistently increased their time spent as the session progressed; Pattern 2 : participants (n=6) gradually increased their time per topic before a slight decline near the end; Pattern 3: participants (n=2) maintained relatively even time across topics.}
    \label{fig.time}
\end{figure}

To analyze how participants actively participate in reminiscence activity in RemVerse, we analyzed the time that each user spent on each topic, and how the normalized time changed in the experience process. In general, most of the participants (N = 12) tend to talk more about their memories as they progressed in RemVerse (Fig. \ref{fig.time}. a, b), only two participants spent approximately even time on each topic (Fig. \ref{fig.time}. c). Furthermore, we observed that most participants (N=13) began to actively share their memories without any prompts from the agent at a certain point at the later stage. As a result, the turn-taking between the participant and the agent between the two stages decreased significantly. 

\subsubsection{RemVerse assisted participants’ reminiscence to become more active and self-initiated.} Based on the results of user-study, we analyzed participants' reminiscence behavior through the turn-taking number with the agent, the contents of their shared memories, and their self-explanation from semi-structured interview. Consequently we observed that participants became increasingly engaged and self-driven in their reminiscence as the session progressed.

\begin{figure}[ht]
    \centering
    \includegraphics[width=0.5\textwidth]{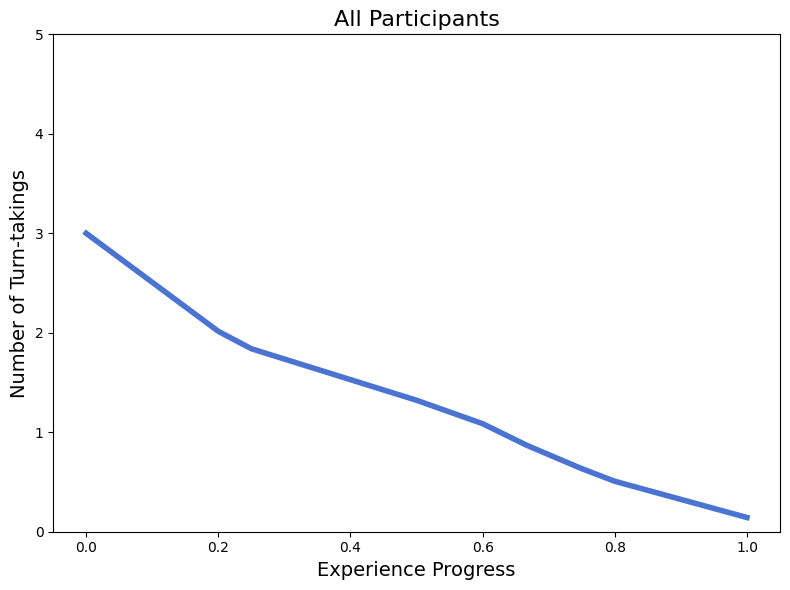}
    \caption{\textit{Mean Turn-taking Number} - \textit{Experience Progress}. As is shown in the figure, the \textit{Mean Turn-taking Number} decreased as participants progressed in RemVerse} 
    \label{fig.turn}
\end{figure}

First, \textbf{as participants progressed in RemVerse, they began to recall and share their recollections more active and spontaneously.} Quantitatively, we observed that for all participants (N=14), the number of turn-takings between the agent and participants decreased over the course of the session (Fig. \ref{fig.turn}), while the length and depth of participants' narratives increased. For instance, P9 initially needed three exchanges with the agent to share a story about her old bicycle—responding to prompts like \textit{``... Did you often ride it as a child?''} and \textit{``Did anything interesting happen after you fell?''}. However, in the later stage of her session (from 60\% to the end), she spoke at length with few assistance, and the agent only contributed once. \revision{Also, this trend was echoed in participants' feedback. Many (N=8) noted that while they initially relied on the agent for support, they gradually found themselves recalling memories more freely as they engaged with the immersive environment. For example, P11 remarked, \textit{``At first, I needed the agent to help me get started, but as I walked around and saw the old things, the memories just started coming back. I didn’t need any more prompts.”} Similarly, P5 reflected, \textit{“I felt like the agent was a guide, but after a while, I was able to tell the stories without needing help. It felt like I was in my own memories, just remembering on my own.”}} 

\revision{In addition, several participants (N=5) even chose to re-visit previously explored parts of the environment in the middle of the session or after explored the whole street. This behavior was not prompted by the agent but initiated by the participants themselves. For example, P3 returned to a front door near the street entrance and remarked: \textit{``Just now at the corner, I saw that convenience store—it reminded me that after school, I used to linger by the small shop near our alley to buy candy. Then I would come home, and my family would be waiting for me, smiling, with dinner... just like this one...”} In a follow-up interview, she explained: \textit{``I passed this earlier and I felt like I was walking through a real scene from the past. It wasn’t until the convenience store reminded me of my childhood buying snacks that I realized this house front was actually part of my memory.”} It also echos with the goal of RemVerse: rather than controlling the flow, RemVerse fosters open-ended exploration where users feel empowered to circle back and deepen prior recollections.}

Also, \textbf{we discovered that participants tended to spend more time on topics} as they progressed in RemVerse (Fig.\ref{fig.time}. a,b). Although some participants (N=6) spent less time in the final stage (they self-reported that they felt little bit tired but still wanted to explore the environment), they still spent more time on topics, and they self-reported that they want to continue to share their memories if there was still space in RemVerse, as is shown in Fig. \ref{fig.time}. a.

\revision{Together, these findings suggest that participants began to actively navigate and recall their own memories, supported but not directed by RemVerse’s interactive features.} While the participants needed prompts and cues from early reminiscence, they took initiative as they became immersed in the environment and engaged with generative tools. The decreasing number of agent-user turn-takings, combined with richer, longer narratives and self-initiated behaviors like re-visiting previous locations or sharing memories without prompts, demonstrates a growing sense of autonomy and confidence. 


\subsubsection{RemVerse triggers and concretizes user's reminiscence through collective Interactions.} Participants unanimously reported that RemVerse provided a realistic and engaging experience that transported them back to their youth. Specifically, they noted that the immersive old-style environment created the illusion of returning to their former living spaces; the generative functions helped materialize their memories; and the agent guided them step-by-step to unfold and structure their recollections. \revision{These elements did not operate in isolation but worked in concert, with each feature co-worked with the others to support a rich, layered reminiscence process.}

First, \textbf{RemVerse triggers participants' reminiscence and renders those memories accessible by offering users the opportunity not only to watch but also to explore a virtual environment resembling the places they once lived, which may have faded from memory or are no longer accessible through videos or photos.} At the start of the user study session, most participants (N = 12) self-reported that they could not clearly remember the past. When attempting to talk about their experiences without assistance or only through photos with friends, they often forgot details or could only recall limited aspects, leading to very superficial conversations. However, during the VR session, all participants not only recalled their detailed past experiences but also organized these memories into relatively comprehensive stories when communicating with the agent. \revision{For example, P11 shared two long stories at the first sight of the pre-set sewing machine in the front door of a house.}, one of the stories was as follows: \textit{`` On the left side, there was a sewing machine. We still have two sewing machines at home, which brings back a lot of memories. It was the 1970s, and I was about 10 years old...". In semi-structured interview, she reported that in seeing the sewing machine outside of the house, she suddenly remembered the similar scene when her aunt }  

\revision{\textbf{Moreover, the agent, immersive environment, and generative models work collaboratively to turn abstract memories into concrete visualizations, forming a dynamic loop that supports deepening recollection.} 
When participants were triggered by the environment, and described objects or scenes that they remembered and then paused, the agent often followed up by inviting them to generate a visual representation (\textit{e.g., ``You’ve mentioned your bicycle’s look, can you show me your old bicycle?''}), and even refine their sometimes vague description to support them generate their expected visual cues. This moment of co-creation deepened participants’ engagement with the system and their own memories. A particularly illustrative example is P5, who first recalled childhood experiences with friends outside a convenience store while reminiscing about snacks. Triggered by the store’s visual context and a prompt from the agent, she re-visited another memory involving riding a bicycle with friends. She then actively re-generated a bicycle to aid this recollection and continued: \textit{``We used to ride together to go see movies...''}. This behavior emerged from the interplay of three components: the evocative VR environment, the agent’s timely intervention, and the user-driven use of generation tools. Similar sequences were observed across participants, indicating that this synergy was not an isolated incident but a behavior that supported fluid and layered memory recall. Importantly, this interactive loop between the agent’s prompts, the immersive environmental cues, and generative visuals was observed across nearly all participants. All participants (N=14) used the generation functions at least twice. In several cases (N=17), image or object generation was initiated directly by the agent’s prompt (\textit{e.g., ``Can you show me the scene of the old vegetable stall that you just mentioned?''}), while in others (N=14), participants took initiative themselves to support their ongoing storytelling.

In several cases (e.g., P2, P6), participants even corrected and re-prompted the generated images with the help of the agent, and accessed further layers of memory through this iterative refinement process. This cyclical loop consists of verbal recollection, system response, and user adjustment demonstrates how the system jointly scaffolded memory reconstruction, and supported user themselves to actively recall their past. For instance, after generating a new basket image, \textit{P2 smiled and said, ``This basket looks more like the ones we used back then.''}, in conversation with the agent, he then elaborated, saying, \textit{``Back then, we didn’t just carry baskets in our hands—we also used shoulder poles to carry them,''} while physically mimicking the motion of lifting and balancing a pole across his shoulders.

This ``loop'' refers to a recurring interaction behavior: the participant is inspired by some cues and recalls part of the memory; the system scaffolds it through generated new visual cues and prompt; this new input triggers further elaboration or correction from the participant, and even lead to new generations. Each step helped participants to further recall their memory. 
Combined with contextual cues from the environment and prompts from the agent, this scaffolding allowed participants to re-enter the emotional, social, and physical dimensions of their lived past. The result is a multi-sensory, affect-rich, and performative reminiscence experience.}

\subsubsection{RemVerse deepens participants' memories through the step-by-step guidance of the agent.} \label{deepen}Generally the interaction between agent and participants in RemVerse could be classified into three categories: initiate, unfold, and evoke.

First, \textbf{the agent helped participants initiate their memory when they noticed an element but hesitated.} Most participants (N=12) reported that the agent played a key role in helping them initiate their reminiscence activity, especially when they were uncertain where to begin. For example, P11 hesitated after seeing the bicycle, and in the interview she explained \textit{``I saw the old bicycle, but I did not know where to begin with."}, then the agent started the topic with the prompt \textit{``...It must carry many memories. Is there any particular moment or trip that stands out..."}, as is shown in Fig. \ref{fig.example}. This kind of initial prompt helped participants overcome the hesitation and sparked their memories, and participants could start sharing their memory (\textbf{e.g.} P9 responded\textit{``When I saw this bicycle, I immediately thought of the Phoenix brand bicycle that we really wanted and loved back then..."}).

Second, \textbf{the agent helped participants further unfold their memories, guiding them to actively extend simple descriptions into memorial stories filled with details and emotions by themselves.} Referring to our coding results, nearly all participants (N=13) only stayed at a shallow level of reminiscence at first, they simply described the appearance of the object in their memory, or how they used the object. For instance, when sharing the memory related to the bamboo basket, P3 firstly only simply described it.\textit{``Bamboo baskets were commonly seen in our lives in the past. Every family would have various sizes of bamboo baskets, used to hold different items."}) However, then the agent prompted \textit{``...The weaving craftsmanship of these baskets was especially delicate, making them perfect for holding food, miscellaneous items, and even for small crafts. Did your family have many bamboo baskets like this in the past? ..."}, the user shared a very fluent and comprehensive story related to bamboo basket \textit{``When I was a child, our family mainly ate flour-based foods... After steaming, since we didn’t have the modern freezing conditions, we would store the food in bamboo baskets to prevent rats...  So, what we did was put a few hooks on the beams or rafters in the room, and hang the steamed buns and other foods up high..."}. The agent's prompt encouraged P3 to elaborate, transforming the initial, superficial recollection into a richer, more vivid narrative. The similar behavior is also shown in Fig. \ref{fig.example}. 2. Similarly, many participants (N=12) expressed how the agent's guidance unfolded their memories from simple descriptions to deeper recollections filled with emotional significance. As participants received prompts like \textit{``Did you often ride it on trips or for leisure?"} or \textit{"What was special about the time you spent with this object?"}, they were able to connect seemingly insignificant memories into a  more intricate and emotionally charged narrative.

What's more, \textbf{the agent helped evoke deep memories that participants may not have otherwise thought of, including those they believed to be forgotten.} Some participants (N=7) reported that RemVerse evoked their memories, for example, P6 self-reported that the agent helped her recall the memories of her late father through interactions with the agent. She explained, \textit{"Through the conversation with the little boy, my memories were awakened. My father has been gone for a  long time, but through the conversation with the boy, I realized that I still clearly remembered scenes of living and talking with my father, I initially thought the memories were faded."} Those deep memories buried in their mind were even irrelevant to the initial triggering topics, with the step-by- step guidance of the agent, they could find their precious but long-buried memories. 

\subsection{Participants’ Feedback and Expectation for RemVerse}

\begin{figure}[ht]
    \centering
    \includegraphics[width=0.9\textwidth]{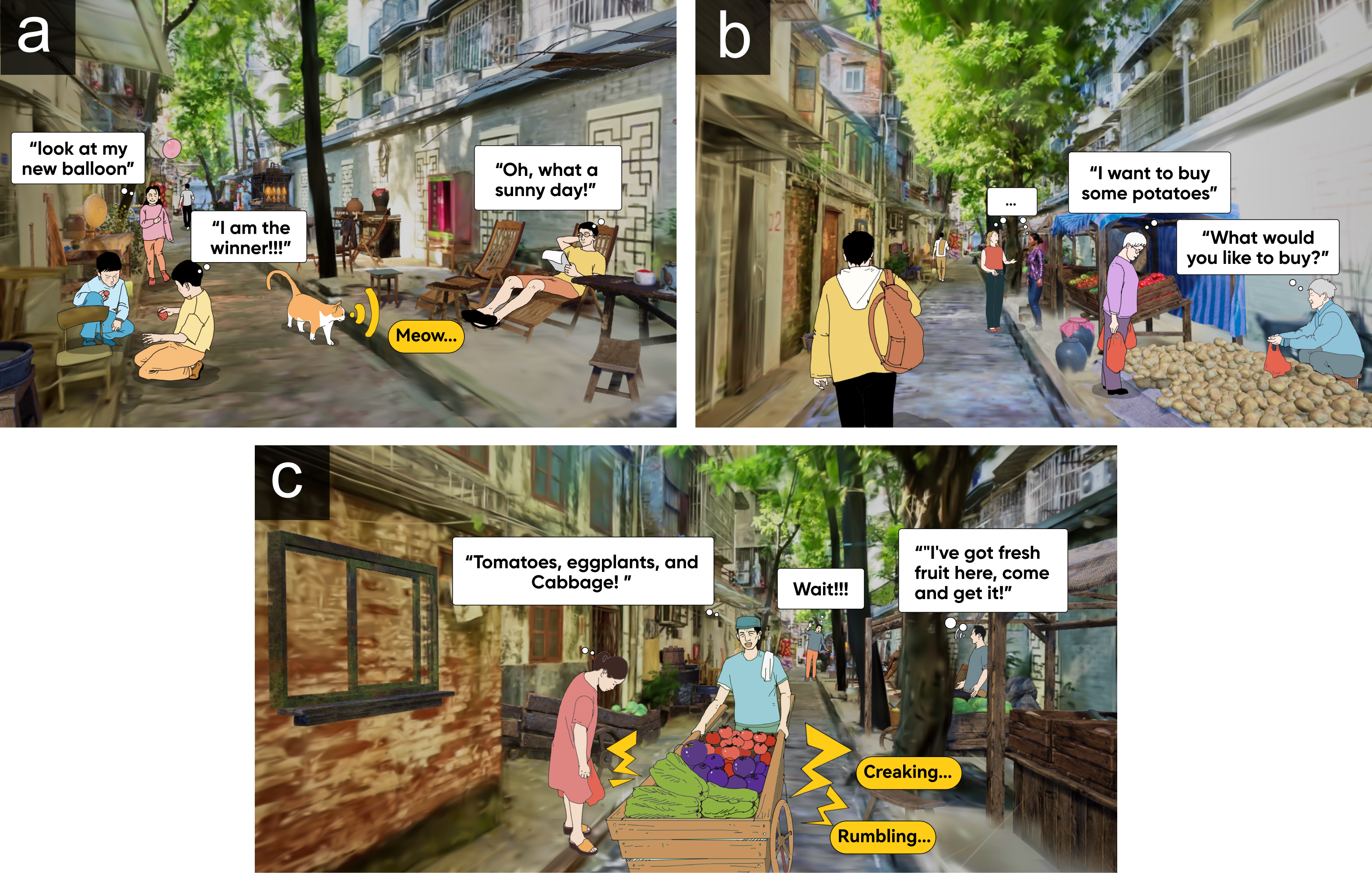}
    \caption{Participants' Synthesized Feedback. We gathered participants' sketches, together with their verbal feedback, we synthesized the suggestions into three scenes.}
    \label{fig.interview}
\end{figure}

Except from our observation, analyze, and summary of semi-structured interview, participants also reported several valuable suggestions, for which we assisted them to visualize in the semi-structured interview session. We synthesized three points from their verbal feedback, and three scenes from their sketches, as is shown in Fig. \ref{fig.interview}. Also, based on their behavior and semi-structured interview, we compared their preference and usage of functions.

\subsubsection{Participants Expect a More Dynamic Environment} 
Although we set the NPCs in the environment(Fig. \ref{fig.remverse}. a), some participants (N = 4) still felt that the virtual scene lacked sufficient dynamic elements, such as moving pedestrians, children, and pets, after we assisted them to sketch together and synthesized their visual feedback, we get the scene of Fig. \ref{fig.interview}. a. P7 and P8 reported that \textit{``...I noticed the NPC, although they moved a little, I still felt there shall be other pedestrians and children.}  This highlights the need for future designers to consider incorporating richer, more responsive social and environmental cues, and create a more vibrant, lifelike setting that resonates with users’ memories of past social contexts. 

\subsubsection{Participants Expect a More Sound-rich Environment} 
We have designed various sound effect in the environment apart from the agent, as is shown in Fig. \ref{fig.remverse}. a. However, some participants (N=3) reported that they expect more voice in the environment. {``I heard the vendor's hawking in the environment, that's the voice of my childhood... but I also expect there could be voice from..."}. They mainly focused on two parts: the voice of the interaction of various NPCs, like pedestrians talking with the vendor (Fig. \ref{fig.interview}. b), and the voice from pets and tools (one mentioned and drew the voice of the push cart, as is shown in Fig. \ref{fig.interview}. a, c).

\subsubsection{Accessibility of Interaction Mechanisms} 

Apart from the virtual environment, some participants (N=3) also reported that VR controllers were difficult to use, pointing to potential usability barriers. This suggests a need for more accessible interaction designs, possibly through simplified controllers, voice commands, or gesture-based interfaces to ensure that older adults can comfortably navigate and interact with the VR environment without undue physical strain or cognitive load.




\section{Discussion}
In this study, we identified how AI-assisted VR environments helped older adults perform reminiscence activity both qualitatively and quantitatively. In this section, we first discussed key takeaways, and then we proposed key takeaways and design implications for future immersive reminiscent systems.

\subsection{Key takeaways}

\textbf{RemVerse assisted reminiscence through immersive, explorable environments with rich cues.} RemVerse helped participants with their reminiscence activity by immersing them in an explorable environment that closely resembled places they had once lived in the past, and it evokes a wide range of personalized memories through the rich visual information in the environment. Unlike previous reminiscence activities \cite{cuevas2020reminiscence}, in which participants followed the cues from photos or videos, RemVerse provided participants with explorable visual cues (environment, old objects, NPCs...) , sound, in an explorable way. And they could visualize their memories through generative functions. These abundant visual cues encouraged participants to pause in front of objects or scenes that triggered their memories and interests, prompting them to actively reflect, recall, and share their experiences. 

\textbf{RemVerse assisted reminiscence by enriching self-expression.} We found that the generative functions assisted participants by transforming their verbal descriptions into visual representations, which aligns with previous work showing that generative models can support older adults' reminiscence by enhancing their expression \cite{chenlimusic}. Especially, our findings extend this understanding by demonstrating that generative models combined with the agent in-turn facilitates the elaboration of memories. Even when the generated content is not entirely accurate, it still promotes reminiscence, since participants engaged in correcting the generated content and even re-generate the image, which further stimulated their recollections.

\revision{\textbf{RemVerse facilitated the deepening of participants' memories.}
We analyzed how the agent co-worked with the 3D environment and generative models to support layered memory recall, and observed a recurring interactive loop: the participant is inspired by certain environmental or conversational cues and recalls part of a memory; the system scaffolds this recollection by generating visual representations or offering further prompts; this new input then triggers the participant to elaborate in more detail, refine their story, or even correct and re-generate the content. The process often leads to deeper and more emotionally rich recollections. Moreover, we identified three core roles that the agent played in this process: \textit{initiating} topics when participants hesitated, \textit{unfolding} memories by encouraging elaboration beyond superficial descriptions, and \textit{evoking} long-buried or marginalized memories through affective prompts and contextual cues. These findings align with previous studies \cite{QUhuaminpoto, quizrem} showing that conversational agents can foster engagement and narrative construction. However, our study further demonstrates how the agent’s effectiveness is amplified when embedded within a richly explorable environment and combined with generative tools, allowing users to not only verbalize but also visualize and embody their memories.}

\textbf{RemVerse assisted reminiscence by encouraging active, user-led exploration}  
As participants progressed through the RemVerse experience, they became increasingly active and autonomous in their memory recall. In our study, we not only confirmed this trend quantitatively—by observing a general decrease in agent-participant turn-taking and an increase in narrative length—but also extended it through qualitative evidence. Participants not only spoke more, but also engaged more meaningfully in both generating visual contents and locomotion behavior. In experiment, many participants transitioned from responding cues and prompts to freely sharing their memories with little or no assistance. In several cases, participants even revisited earlier scenes on their own to continue or deepen a memory thread. These behaviors indicate a shift from system-led to self-initiated interaction, suggesting that RemVerse supports more than passive engagement, it also enables sustained, self-directed participation in the reminiscence process.

\subsection{Design Implications}
We synthesized the feedback from participants, and based on our observation and analysis, we propose design implications as below.

\subsubsection{\textbf{Promote empathetic and contextually aware interaction}}
\revision{As is shown in the paper, participants' behavior of performing reminiscence activities changes during the time, it not only includes their tendency to self-initiate the topics in later stages, but also related to different context and emotion along with different memories.} Namely, during the study, participants mentioned that the agent's tone was too \textit{``light"}, making some deeply emotional and heavy memories sound overly casual and cheerful, and their emotions tied to these memories are much deeper and more profound. Thus, the fact led us to consider \textbf{using AI to understand users' personalized memories, their context, and emotions, to better promote reminiscence activity}, as these factors reflect users' higher-level needs in expressing their memories. Based on our findings, our system has partially addressed these aspects by facilitating memory recall and enabling users to share more detailed narratives. However, it still cannot fully understand and accurately respond to participants' various emotions.

First, RemVerse could be improved to empathetically understand the emotions tied to users' memories and uncover the complex, underlying feelings that may not be immediately obvious. By incorporating more advanced emotion-detection mechanisms \cite{LLMs&Sentiment, HumanRobotLLM, conversationagent}, the agent could not only recognize explicit emotional cues but also detect nuanced emotional states, such as nostalgia, regret, or unresolved feelings, that users may experience while recalling their memories.
Moreover, emotional responsiveness could be further refined by dynamically adapting the agent’s behavior to the user’s level of engagement. For instance, if the user appears less engaged or uncertain, the agent could take a more proactive, guiding role to help initiate or sustain the memory recall process. Conversely, when users are highly engaged and self-initiated in their storytelling, the agent could shift to a more passive, listener-oriented stance, creating space for uninterrupted reflection. Although this adaptive conversational strategy was already partially implemented in our current system: as users became more active over time, the agent naturally reduced its interventions, encouraging greater autonomy, future work could build on this by making such adaptability more intentional and emotion-aware.

\revision{Second, RemVerse could be refined to create a contextually aware adaptive environment that evolves in response to the user's interactions. Here ``adaptive'' refers to both context, behavior, and physical condition. Based on the context of participants, and the engagement with previous memories and behaviors, the environment could dynamically adjust to better align with the user's emotional and cognitive state. For example, the scenery, sounds, or objects in the virtual space could subtly change to support deeper engagement or provide a comforting atmosphere as the user revisits more complex memories. This adaptability could also involve modifying the pacing and structure of reminiscence activities, allowing the environment not only to respond to  the depth and intensity of the user’s emotions and memory recall, but also to adapt to their body energy and physical condition.}

\subsubsection{\textbf{Enable Flexible Edition over Users' Personal Memories}}

As is mentioned in result, the generated content in RemVerse may not always be completely accurate, yet they still play a crucial role in triggering users' memories and facilitating the elaboration of those memories. In fact, most participants still exhibited a high level of engagement when the generated objects or images did not match their recollections. Rather than being frustrated by the inaccuracies, they showed an active interest in correcting them, engaging deeply with the content in order to make it more reflective of their own memories. This process of "editing" the generated elements became an integral part of their reminiscence experience.

Interestingly, many participants expressed a sense of self-satisfaction when they were describing corrections to the generated content, aiming to make it more closely align with their personal recollections. This act of verbal correction was not just seen as a simple fix but as a meaningful and empowering engagement with their memories. Although the ability to physically edit the virtual elements is not yet available in the current version of RemVerse, we suspect that participants would greatly enjoy the opportunity to take control of the virtual environment in the future, shaping it to reflect their unique past experiences. At present, this verbal engagement served to deepen their reflection and foster a sense of ownership and agency over their memories. This, in turn, enhanced their emotional connection to the reminiscence process. 

In the future, \textbf{providing participants with the ability to directly edit the generated content could offer them a more active role in shaping their reminiscence experience.} While this feature is not yet available in RemVerse, we speculate that allowing users to modify the virtual elements could further enhance their sense of control over the environment. By enabling them to adjust the generated scenes to better align with their memories, participants might experience a deeper connection to their past. This greater control could potentially encourage more meaningful interactions, allowing users to personalize the experience and engage with their memories in a more dynamic way. We believe that such a feature could increase emotional engagement, as participants would feel more empowered to reconstruct their memories in a way that reflects their individual experiences.

\subsubsection{\revision{\textbf{Tailoring generated content to memory contexts}}}

\revision{Throughout the study, we observed that participants used the object generation function more frequently than the image generation function. While both tools were helpful in supporting memory expression, participants gravitated more often toward the 3D object creation feature.

This usage pattern may reflect a preference for concrete, manipulable representations in a spatial context, especially when participants are immersed in a 3D environment. Unlike unconstrained images, which sometimes deviated from what participants envisioned, constrained 3D object generation offered a more grounded and tangible anchor for recalling and elaborating on memories. This distinction highlights the need for generative models that are sensitive not only to language input, but also to the cognitive and emotional characteristics of autobiographical memory.

Looking forward, this points to a design opportunity: generative systems for reminiscence could be further improved by adapting outputs to better match users’ internal memory context. For example, a fine-tuned memory-sensitive visual model that conditioned on background data collected through agent interaction could generate visualizations that better reflect what the user remembers but may not be able to describe precisely. This could involve constraining the style, material, or structure of the output to fit the temporal and cultural context of the memory being recalled. Such models could help bridge the gap between fuzzy memory fragments and vivid visual representations, deepening engagement and emotional resonance.}

Moreover, transforming static images into videos may provide a more dynamic and immersive experience, which might also be able to provide users with sufficient cues. By animating key elements of the image, such as adding motion or transitions, users could engage with the memory in a more lifelike manner\cite{img2video1, img2video2}. This would create a more immersive environment that could evoke stronger emotional connections, as the dynamic nature of video could better mirror the fluidity of memory itself. Moreover, enabling the conversion of 3D images into fully interactive 3D models could allow participants to directly interact with the generated objects or scenes in future versions of RemVerse. By enhancing the interactivity of both images and models, we could create a more seamless and engaging experience, allowing users to not only view their memories in a static form but also engage with them in real-time \cite{img3D}. This would further deepen participants' emotional connections to their past experiences, offering a richer and more immersive experience.

\subsubsection{\revision{\textbf{Exploring the impact of personal preference and novelty-induced engagement patterns}}}

\revision{While our study revealed that participants became self-initiated and active in their reminiscence as they progressed through RemVerse, this increase in engagement may be influenced by several factors. 

First, for different participants, some objects may naturally evoke stronger emotional or autobiographical responses than others, leading to deeper engagement regardless of when or where they appear. Certain objects might serve as more powerful memory triggers due to personal relevance or cultural significance. Future research could more systematically examine how different types of cues (e.g., familiar objects, culturally specific symbols, or personally significant locations) affect engagement levels, and how to adapt content delivery to individual memory patterns.

Second, our current study employed an integrated system combining an AI agent, immersive VR environments, and generative content. While this multi-modal approach successfully supported deeper and more autonomous storytelling, it's also important to isolate the contribution of each modality. To address this, future studies will include controlled comparisons with alternative reminiscence methods such as traditional photo-based sessions, AI-only conversational systems, and standalone VR environments without AI guidance. By comparing these approaches, we could to better understand the relative strengths of each modality, and evaluate whether the collective system offers unique advantages in memory recall and emotional engagement.

Third, to further disentangle the effects of individual system components, future work will involve component-isolation experiments. These studies will systematically manipulate the presence or absence of core elements—AI guidance, immersive VR, and generative media—to examine how each contributes to the reminiscence experience. For instance, we plan to compare: (1) AI-only interaction in a neutral environment, (2) VR exploration without conversational agents, (3) generative content without immersion or conversation, and (4) the full system. This would provide a more fine-grained understanding of how each component supports memory retrieval, narrative construction, and emotional depth. Such research could also inform adaptive system designs that dynamically emphasize or de-emphasize components based on user needs.

To gain a more comprehensive understanding of how different system components and user factors shape the reminiscence experience, future studies will integrate behavioral data, physiological signals (e.g., heart rate variability, skin conductance, fNIRs), and validated reminiscence engagement scales. This multi-dimensional evaluation framework will enable a deeper analysis of emotional engagement, cognitive effort, and user-state transitions during interaction. Furthermore, we plan to conduct these studies in larger-scale, more diverse virtual environments(e.g., virtual neighborhoods, homes, or public spaces), and over longer deployment periods. These extended and context-rich settings will allow us to examine how engagement patterns develop over time, how sustained interaction influences memory expression, and how novelty, personal relevance, and design features dynamically interact across sessions.}

\section{Conclusion}
In this study, we introduced RemVerse, an AI-assisted VR prototype designed to enhance reminiscence activities for older adults by recreating historical urban streetscapes in an interactive 3D environment. By integrating 3DGS for realistic and efficient environmental reconstruction, generative models for personal visual content creation, and AI-driven communication agents, RemVerse provided a immersive and personalized experience for older adults, supported their reminiscence activity. The results from the user study involving 14 participants highlighted the effectiveness of RemVerse in triggering, enriching, and evoking memories, with the interactive elements and AI agents playing a significant role in promoting deeper reminiscence. Participants expressed positive feedback when interacting with familiar objects and scenes from their past, demonstrating the potential of VR and AI in supporting reminiscence activity. Moreover, the semi-structured interview revealed several areas for improvement, including enhancing the system’s accessibility and tailoring the experience to individual preferences. Based on these findings, we also offered design implications for future AI-assisted VR systems, emphasizing the importance of personalization, interactivity, and edition over generated content. As we move forward, we hope that RemVerse and similar systems can be valuable, efficient, and convient tools for promoting mental and emotional well-being for older adults, enriching their reminiscence activities, and supporting their connection to the past in an increasingly fast-paced world.
\section{Acknowledgment}
This work is partially supported by Guangzhou-HKUST(GZ) Joint Funding Project (No. 2024A03J0617), Guangzhou Higher Education Teaching Quality and Teaching Reform Project (No. 2024YBJG070),  Education Bureau of Guangzhou Municipality Funding Project  (No. 2024312152), Guangdong Provincial Key Lab of Integrated Communication, Sensing and Computation for Ubiquitous Internet of Things (No. 2023B1212010007), the Project of DEGP (No.2023KCXTD042), and Artificial Intelligence Research and Learning Base of Urban Culture (No. 2023WZJD008).

\bibliographystyle{ACM-Reference-Format}
\bibliography{reference.bib}
\end{document}